\documentclass[12pt]{article}
\newcommand{\beq}{\begin{equation}}
\newcommand{\eeq}{\end{equation}}
\newcommand{\beqa}{\begin{eqnarray}}
\newcommand{\eeqa}{\end{eqnarray}}
\newcommand{\beqar}{\begin{eqnarray*}}
\newcommand{\eeqar}{\end{eqnarray*}}

\newcommand{\eps}{\epsilon}

\newcommand{\inn}{\!\cdot\!}

\newcommand{\la}{\lambda}

\newcommand{\z}{\zeta}

\newcommand{\eg}{{\it e.g.,}\ }
\newcommand{\ie}{{\it i.e.,}\ }
\newcommand{\labell}[1]{\label{#1}} 
\newcommand{\reef}[1]{(\ref{#1})}

\newcommand\veps{\varepsilon}

\newcommand\cF{{\cal F}}
\newcommand\cN{{\cal N}}

\newcommand\cM{{\cal M}}

\newcommand\cB{{\cal B}}

\newcommand\tq{{\tilde q}}

\begin{document}

\vspace*{1cm}

\begin{center}
{\bf \Large  S-duality of  S-matrix }

\vspace*{1cm}

{Mohammad R. Garousi\footnote{garousi@ferdowsi.um.ac.ir} }\\
\vspace*{1cm}
{ Department of Physics, Ferdowsi University of Mashhad,\\ P.O. Box 1436, Mashhad, Iran}
\\
\vspace{2cm}

\end{center}

\begin{abstract}
\baselineskip=18pt

Recently, it has been observed that  the kinematic factor of the disk-level S-matrix element of two RR two-forms and  the disk-level S-matrix element of two B-fields on the world volume of D$_3$-brane are compatible with the standard rules of S-duality. Inspired by this observation, we speculate that the S-matrix elements on  the world volume of D$_3$-brane  are invariant under the $SL(2,Z)$ transformation. Compatibility with S-duality requires  the S-matrix elements on the world volume of D$_1$-brane and D$_5$-brane to be extended to the $SL(2,Z)$-covariant form.  In particular, this implies  
the  S-matrix elements on the world volume of F$_1$-string and NS$_5$-brane at strong coupling to be related by S-duality to  the disk-level S-matrix elements  of D$_1$-string and D$_5$-brane, respectively. 
The contact terms of these S-matrix elements at order $O(\alpha'^0)$  produce   a  Born-Infeld and Chern-Simons type effective action  for both F$_1$-string and NS$_5$-brane. They are  consistent with the $SL(2,R)$-covariant action appears in the literature.

\end{abstract}

\vfill
\setcounter{page}{0}
\setcounter{footnote}{0}
\newpage


\section{Introduction } \label{intro}
It is known that the  type II superstring theory  is invariant under  T-duality \cite{Kikkawa:1984cp,TB,Giveon:1994fu,Alvarez:1994dn,Becker:2007zj}  and  S-duality \cite{Font:1990gx,Sen:1994fa,Rey:1989xj,Sen:1994yi,Schwarz:1993cr,Hull:1994ys,Becker:2007zj}.
Compatibility of a given solution of equations of motion with these dualities can be used to generate new solutions \cite{Hassan:1991mq,Cvetic:1995bj,Breckenridge:1996tt,Costa:1996zd}. In this paper, we would like to apply this  compatibility to the other on-shell quantities, \ie the S-matrix elements.

The S-matrix elements of any  nonlinear gauge theory must satisfy the Ward identity.   This is the linearized gauge transformations in the momentum space. For example, the S-matrix elements of a gravity theory must be invariant under the replacement $\veps_{\mu\nu}\rightarrow \veps_{\mu\nu}+p_{\mu}\z_{\nu}+p_{\nu}\z_{\mu}$ where $\veps_{\mu\nu}$ is the polarization of external graviton, $\z_{\mu}$ is an arbitrary vector  and $p\inn\z=0$. 
We expect  similar Ward  identities for the global  S and T dualities in string theory. Since there is no derivative in the duality transformations,  one should replace a polarization tensor with its dual  tensor, which is related to the original one  by the linear  duality transformations.  This replacement may  produce  new S-matrix elements which are related to the original one by the linear dualities. On the other hand,  we expect the S-matrix elements to be invariant/covariant under the nonlinear duality transformation on the background fields in the S-matrix element. 

The T-duality holds order by order in string loop expansion \cite{Becker:2007zj}. This indicates that  a given S-matrix element at any loop order is invariant/covariant under the T-duality transformation on the background fields. The invariance of the S-matrix element  under linear T-duality transformation on the quantum states then can be used to  extend  the S-matrix element to  a family of   S-matrix elements.   We call such S-matrix elements the  T-dual S-matrix multiplet.  Let us examine this idea for the  simple example of  disk level S-matrix element of  two gravitons.  In general, the disk-level two-point function of closed strings is given by \cite{Garousi:1996ad,Hashimoto:1996bf}
\beqa
 A(D_p;\veps_1,\veps_2)&\sim&T_p\alpha'^2K(D_p;\veps_1,\veps_2)\frac{\Gamma(- t/4)\Gamma(- s)}{\Gamma(1-t/4- s)}\delta^{p+1}(p_1^a+p_2^a)\labell{A}
 \eeqa
 where $\veps_1,\, \veps_2$ are the polarization of external states. In this amplitude, the Mandelstam variable  $s=-\alpha'(p_1)_a(p_1)_b\eta^{ab}$ is the momentum flowing along the world volume of brane, and  $t=-\alpha'(p_1+p_2)^2$ is the momentum transfer in the transverse directions on the brane.   The background  metric $\eta_{\mu\nu}$ is the string frame metric\footnote{Our index convention is that the Greek letters  $(\mu,\nu,\cdots)$ are  the indices of the space-time coordinates, the Latin letters $(a,d,c,\cdots)$ are the world-volume indices and the letters $(i,j,k,\cdots)$ are the normal bundle indices.}. The $s$-channel describes the open string excitation of the D-brane, and the $t$-channel describes the closed string couplings to the D-brane. The background has no  axion,  hence, the  tension of the intermediate string in the string frame is $T_{F_1}=1/(2\pi\alpha')$. In above equation, $T_p$ is the tension of external D$_p$-brane in the string frame, \ie
 \beqa
 T_p&=&\frac{1}{g_s(2\pi)^p(\alpha')^{(p+1)/2}}\labell{tension}
 \eeqa
 where $g_s=e^{\phi_0}$ is the  closed string coupling.  We have normalized this amplitude and all other amplitudes in this paper by requiring them to be consistent with T-duality. We will not fix, however, the numeric factor of the amplitudes.
 
 The kinematic factor in \reef{A} for two gravitons is a lengthy expression in terms of the graviton polarizations  \cite{Garousi:1996ad}. Using on-shell relations, this factor can be  written  in terms of the linearized curvature tensor of the external states in the momentum space as \cite{Bachas:1999um}
\beqa
K(D_p;h_1,h_2)&=&R_{1abcd}R_2^{abcd}-2\hat{R}_{1ab}\hat{R}_2^{ab}-R_{1abij}R_2^{abij}+2\hat{R}_{1ij}\hat{R}_2^{ij}\labell{DBI}
\eeqa
 The indices are raised and lowered by the flat metrics $\eta_{ab}$ and $\eta_{ij}$. In above equation $\hat{R_1}_{ab}=\eta^{cd}{R_1}_{cadb}$ and $\hat{R_1}_{ij}=\eta^{cd}{R_1}_{cidj}$. The linearized curvature tensor corresponding to the first graviton is
\beqa
{R_1}_{\mu\nu\rho\la}&=&\frac{1}{2}({h_1}_{\mu\la,\nu\rho}+{h_1}_{\nu\rho,\mu\la}-{h_1}_{\mu\rho,\nu\la}-{h_1}_{\nu\la,\mu\rho})\nonumber
\eeqa
where the metric in the curvature tensor is $\eta_{\mu\nu}+{h_1}_{\mu\nu}$ and ${h_1}_{\mu\nu}$ is the graviton polarization tensor. The commas denote partial differentiation in the momentum space. 

To apply the T-duality on the 2-point function, one should first use the nonlinear T-duality on the background, and then use linear T-duality on the quantum states. If one implements T-duality along a world volume direction  of D$_p$-brane, then  the background fields transform under the nonlinear T-duality to
\beqa
T_p\delta^{p+1}(p_1^a+p_2^a)&\longrightarrow&T_{p-1}\delta^{p}(p_1^a+p_2^a)\labell{backT}\\
s,t&\longrightarrow&s,t\nonumber
\eeqa
where  we have used  the assumption implicit in the T-duality that fields are independent of the Killing coordinate, \eg $\delta^{p+1}(p_1^a+p_2^a)=\delta^{p}(p_1^a+p_2^a)\delta(0)$ where $\delta(0)=2\pi R$ and $R$ is the radius of the compact direction on which the T-duality is applied. So the D$_p$-brane  of type IIA/IIB transforms to the D$_{p-1}$-brane  of type IIB/IIA. The quantum fluctuations $h_1,\, h_2$,
 in the kinematic factor \reef{DBI} transform to the following expression under the  linear T-duality \cite{Garousi:2009dj}:
\beqa
K(D_{p};B_1,B_2)&=&
\frac{1}{6}H_{1ijk,a}H_2^{ijk,a}+\frac{1}{3}H_{1abc,i}H_2^{abc,i}-\frac{1}{2}H_{1bci,a}H_2^{bci,a}
\labell{LDBI}
\eeqa
 where $B_1$, $B_2$ are the polarization tensors of the antisymmetric B-field and $H_1$, $H_2$ are their field strengths. Using the above transformations, one observes that the T-duality transformation of the disk-level 2-point function of gravitons is given by the amplitude \reef{A} in which the kinematic factor is given by the above expression. It has been shown in  \cite{Garousi:2009dj} that this result is in fact the disk-level 2-point function of B-field. So the disk-level 2-point function of graviton and the the disk-level 2-point function of B-field come together as  a T-dual multiplet.

  Another example that shows the S-matrix elements are invariant under  T-duality is the disk-level S-matrix element of one RR and one NSNS vertex operators  which is given by \reef{A} with appropriate kinematic factor \cite{Garousi:1996ad}. This factor has  been studied in details in \cite{Garousi:2010ki,Garousi:2011fc} for various RR forms and NSNS states. It is given by
\beqa
K(D_p;\veps_1,\veps_2)&=&\eps^{a_0\cdots a_p}\left(\frac{1}{2!(p-1)!}[{ F_1}^{(p)}_{ia_2\cdots a_p,a}{H_2}_{a_0a_1}{}^{a,i}-{ F_1}^{(p)}_{aa_2\cdots a_p,i}{H_2}_{a_0a_1}{}^{i,a}]\right.\nonumber\\
&&\left.\qquad\quad+\frac{2}{p!}[\frac{1}{2!}{ F_1}^{(p+2)}_{ia_1\cdots a_pj,a}({R_2}^a{}_{a_0}{}^{ij}-\frac{1}{p+1}{ F_1}^{(p+2)}_{a_0\cdots a_pj,i}(\hat{R_2}^{ij}-\phi_2\,^{,ij})]\right.\nonumber\\
&&\left.\qquad\quad-\frac{1}{3!(p+1)!}{ F_1}^{(p+4)}_{ia_0\cdots a_pjk,a}{H_2}^{ijk,a}\right)\labell{LTdual}
\eeqa
where    $F_1^{(n)}$ is the field strength polarization tensor of the RR potential,  and $\phi_2$ is the polarization of the dilaton which is one.
The sum of the second term in the first line and the last two terms in the second line form a T-dual multiplet, and the remaining terms  form another T-dual  multiplet. Hence, the 2-point functions are invariant under  T-duality.


On the other hand, the S-duality holds order by order in $\alpha'$ and is nonperturbative in the string loop expansion \cite{Becker:2007zj}. This indicates that a given S-matrix element at tree-level is ${\it not}$ invariant/covariant under the nonlinear S-duality transformations on the background fields.  Hence,  one must include the  loops and the nonperturbative  effects \cite{Green:1997tv} to make it invariant. This is unlike the T-duality, in which the tree-level is invariant under the nonlinear T-duality on the background fields, \eg,  equation \reef{backT}.  For concreteness, let us consider the sphere-level 4-point function of gravitons in type IIB string theory \cite{Green:1981ya,Gross:1986iv}. The background flat metric  in the Mandelstam variables is the string frame metric, hence, the Mandelstam variables are not invariant  under the S-duality. In this case, one has to $\alpha'$-expand the amplitude  to be able to discuss the S-duality of the background fields. At the leading order in $\alpha'$, the S-matrix element is invariant under the S-duality.  At the next leading order in $\alpha'$, the amplitude in the Einstein frame are proportional to $e^{-3\phi_0/2}$ where $\phi_0$ is the dilaton background.  This factor is not invariant under the nonlinear S-duality. It has been shown in \cite{Green:1997tv} that the four-graviton couplings at order $\alpha'^4$, which produce $R^4$ couplings in the spacetime \cite{Green:1981ya,Gross:1986iv,Grisaru:1986dk,Grisaru:1986px}, become invariant under the S-duality when one includes the one-loop and the D-instanton effects. The amplitude  at the higher orders of $\alpha'$, which include the other dilaton factors,  can be extended to the S-duality invariant form by adding the higher loops and the nonperturbative  effects \cite{Green:1999pu,Green:2005ba,Basu:2007ru,Basu:2007ck,Green:2010kv}.

In some cases, the consistency of the quantum states of a given tree-level S-matrix element with the linear S-duality  can be used to find a family of  tree-level S-matrix elements. We call such S-matrix elements the S-dual S-matrix multiplet. For instance, consider the sphere-level 4-point function of B-fields, which may be found by requiring the consistency of the four-graviton amplitude \cite{Green:1981ya,Gross:1986iv} with the linear T-duality.  This amplitude can not  be extended to the S-dual form by adding only  loops and the nonperturbative effects. In this case one should includes the sphere-level 4-point function of the RR two-form as well. In this sense, one may find new tree-level  S-matrix elements by using the compatibility of a given S-matrix element with the S-duality.

In this paper, we would like to examine the compatibility of disk-level S-matrix elements with the S-duality. In the case that a given S-matrix element  represents the scattering  from D$_3$-brane, its consistency  with the S-duality may be used to find its S-dual S-matrix multiplet. All the elements are the disk-level amplitudes on  the world volume of D$_3$-brane.  In the case that a given disk-level S-matrix element represents the scattering from D$_1$-string  or D$_5$-branes, its consistency with the S-duality can be used to  fix the form of the  S-matrix elements of  F$_1$-string or NS$_5$-brane at strong coupling.


The outline of the paper is as follows: We begin in section 2 by studying  the S-duality transformations of  the disk-level 2-point functions representing the scattering  from D$_3$-brane. We show how compatibility of the 2-point function of the B-field with S-duality predicts the form of some other disk-level amplitudes. In section 3, we show that the compatibility of the disk-level S-matrix elements of D$_1$-brane or D$_5$-brane with the S-duality generates the  S-matrix elements of F$_1$-string or NS$_5$-brane at strong coupling. The new S-matrix elements imply  that both F$_1$-string and NS$_5$-brane have open D$_1$-string excitation, as expected. In section 4,  we argue that the contact terms of the new S-matrix elements at order $O(\alpha'^0)$ produce a Born-Infeld and Chern-Simons  type effective action for both F$_1$-string and NS$_5$-brane which are  consistent with the $SL(2,R)$-covariant action proposed in \cite{Bergshoeff:2006gs}.

\section{ S-duality of D$_3$-brane amplitudes}

It is known that the supergravity effective action of type IIB is invariant under the S-duality \cite{Becker:2007zj}. The RR four-form is invariant under this duality. On the other hand,  the D$_3$-brane couples linearly  to the RR four-form, so the D$_3$-brane is also invariant under S-duality. We expect then the S-matrix elements on the world volume of D$_3$-brane to be invariant under the S-duality. Let us begin with 
the following disk-level one-point function in the string frame:
\beqa
A(D_p;C_1^{(p+1)})&\sim& T_{p}\epsilon^{a_0\cdots a_p}C^{(p+1)}_{1a_0\cdots a_p}\delta^{p+1}(p_1^a)\labell{amp1}
\eeqa
where $T_p$  is the  tension of  D$_p$-brane in the string frame \reef{tension}. We have normalized the amplitude by $T_p$ in order to make it invariant under the T-duality. It is easy to verify that T-duality along the brane transforms it to $T_{p-1}\epsilon^{a_0\cdots a_{p-1}}C^{(p)}_{1a_0\cdots a_{p-1}}\delta^{p}(p_1^a)$. With this normalization, it is obvious that the amplitude \reef{amp1} for D$_3$-brane case is invariant under the S-duality. 

Next, consider  the massless NSNS one-point function  which is
\beqa
A(D_p;\veps_1)&\sim& T_{p}(\veps_{1a}{}^{a}-\veps_{1i}{}^i)\delta^{p+1}(p_1^a)\labell{amp2}
\eeqa
The standard S-matrix calculation for NSNS states gives the amplitude in the Einstein frame, so the above amplitude is in the  Einstein frame.  We normalized the amplitude by $T_p$ to make  it  invariant under T-duality.  To verify that the amplitude is invariant under the T-duality, one must consider the combination of the gravion and the dilaton 1-point functions because T-duality maps dilaton to graviton. The dilaton amplitude can be read from \reef{amp2} by using the polarization tensor $\veps_{\mu\nu}=(\eta_{\mu\nu}-p_{\mu}\ell_{\nu}-p_{\nu}\ell_{\mu})/2$ where the auxiliary vector satisfies $\ell\cdot p=1$. The sum of the graviton and the dilaton amplitudes in the Einstein frame becomes
\beqa
A(D_p;1)&\sim&T_p(2h_{1a}{}^a+(p-3)\phi_1)\delta^{p+1}(p_1^a)\labell{amp3}
\eeqa
where we have also used the traceless of the graviton polarization tensor. The expression inside the parentheses becomes $2(h_{1a}{}^a-2\phi_1)$ in the string frame which is invariant under the linear T-duality \cite{Garousi:2009dj}. For the case of D$_3$-brane, the last term in \reef{amp3} is zero and the first term is invariant under the S-duality, since the Einstein frame metric is invariant.

Now consider the disk-level two-point function of one $C^{(4)}$ and one NSNS state. This amplitude in the string frame is given by \reef{A} in which  the appropriate kinematic factor is given in \reef{LTdual}.  In the Einstein frame $g_{\mu\nu}$, which is related to the string frame metric $G_{\mu\nu}$ as  $G_{\mu\nu}=e^{(\phi_0+\phi)/2}g_{\mu\nu}$ where $\phi_0$ is the constant dilaton background and $\phi$ is its quantum fluctuation, the amplitude becomes 
\beqa
 A(D_3;\veps_1,\veps_2)&\sim&T_{D3}\alpha'^2K(D_3;\veps_1,\veps_2)\frac{\Gamma(- te^{-\phi_0/2}/4)\Gamma(- se^{-\phi_0/2})}{\Gamma(1-te^{-\phi_0/2}/4- se^{-\phi_0/2})}\delta^{4}(p_1^a+p_2^a)\labell{AA}
 \eeqa
 where the kinematic factor is
\beqa
K(D_3;C_1^{(4)},h_2)&=&\eps^{a_0\cdots a_3}e^{-\phi_0}\bigg[\frac{1}{2!3!}{F_1}^{(5)}_{ia_1\cdots a_3j,a}{R_2}^a{}_{a_0}{}^{ij}-\frac{1}{4!}{F_1}^{(5)}_{a_0\cdots a_3j,i}\hat{R_2}^{ij}\bigg]
\eeqa
The quantum states $C_1^{(4)}$ and $h_2$ are invariant under the S-duality, however, the S-matrix element is not invariant under the nonlinear S-duality on the dilaton background.
To study the S-duality of the background, we have to $\alpha'$-expand the Gamma functions. This expansion is 
\beqa
\frac{\Gamma(- te^{-\phi_0/2}/4)\Gamma(- se^{-\phi_0/2})}{\Gamma(1-te^{-\phi_0/2}/4- se^{-\phi_0/2})}&=&\frac{4e^{\phi_0}}{st}-\frac{\pi^2}{24}+O(\alpha'^2e^{-\phi_0})\labell{expand}
\eeqa
One can easily observe that the leading term of the amplitude which is $\alpha'^0$ order is invariant under the S-duality. The $\alpha'^2$ order terms has the dilaton factor  $e^{-\phi_0}$ which is not 
invariant under the nonlinear S-duality. The higher order of $\alpha'$ has other dilaton factors. None of them  are  invariant under the S-duality. 

The dilaton and axion transform similarly under the S-duality, hence, one expects each of the dilaton  factors in the above amplitude to  be extended to  a function of both dilaton and axion to be invariant under the S-duality. In this way, one can find the exact  dependence of the amplitude on the background dilaton and axion. Note that the 2-point function \reef{AA} has no axion background. It has been shown in  \cite{Bachas:1999um,Garousi:2011fc} that by  adding the one-loop and the D-instanton effects to the $\alpha'^2$-order terms, which can be done  by replacing $e^{-\phi_0}$ with the regularized non-holomorphi Eisenstein series $E_1(\phi_0,C_0)$, one extends the $\alpha'^2$-order terms  to the S-dual invariant form \cite{Bachas:1999um,Garousi:2011fc}. The higher order of $\alpha'$ terms  require other  Eisenstein series to make them S-duality invariant \cite{Basu:2008gt}. 

Under the S-duality, $C^{(2)}\rightarrow B$ and $B\rightarrow -C^{(2)}$ \cite{Becker:2007zj}. The invariance of the S-matrix elements under the S-duality then  indicates that the one-point function of the RR two form and the one-point function of B-field must be zero, as they are. So we consider the two-point function of these states which is given, in the string frame,  by \reef{A} and by the appropriate kinematic factor in \reef{LTdual}.  In the Einstein frame,  the amplitude is the same as \reef{AA} in which the  kinematic factor is:
\beqa
K(D_3;C_1^{(2)},B_2)&=&
\eps^{a_0\cdots a_3}e^{-\phi_0}\bigg[{ F_1^{(3)}}_{a_2a_3i,a}{H_2}_{a_0a_1}{}^{a,i}-{ F_1^{(3)}}_{a_2 a_3a,i}{H_2}_{a_0a_1}{}^{i,a}\bigg]\labell{LCS1}
\eeqa
The expression inside  the bracket can be written in the following $SL(2,R)$ invariant form \cite{Garousi:2011fc}:
\beqa
K(D_3;C_1^{(2)},B_2)&=&
\eps^{a_0\cdots a_3}e^{-\phi_0}{ \cF^T}_{a_0 a_1a,i}\cN \cF_{a_2a_3}{}^{i,a}\labell{LCS2}
\eeqa
where the $SL(2,R)$ matrix $\cN$ is
\beqa
\cN=\pmatrix{0&1 \cr 
-1&0}
\eeqa
and  $\cF=d{\cal B}$ where ${\cal B}$  is
\beqa
{\cal B}=\pmatrix{B_2 \cr 
C_1^{(2)}}
\eeqa
Hence,  the S-matrix element is invariant under the S-duality on the quantum states, whereas it is not invariant under the nonlinear S-duality on the background.
Using the expansion \reef{expand}, one finds again that the amplitude at order $\alpha'^0$ is invariant under the S-duality. The dilaton factor in the $\alpha'^2$-order terms  should again be replaced by $E_1(\phi_0,C_0)$ to make it   invariant. The   dilaton factors in  all higher order can be extended to S-dual forms by adding appropriate loops and D-instanton effects.
 
Consider  the following standard coupling on the world volume of D$_3$-brane at order $O(\alpha'^0)$:
\beqa
T_{D3}\int C^{(2)}\wedge B\labell{CB}
\eeqa
Using the S-duality transformation $C^{(2)}\rightarrow B$ and $B\rightarrow -C^{(2)}$, one finds that it is not invariant under the S-duality\footnote{One may use the $SL(2,R)$-doublets
$
\tq=\pmatrix{1 \cr 
0}$, $ q=\pmatrix{0 \cr 
-1}
$
and rotate them at the same time that rotate the doublet $\pmatrix{B \cr 
C^{(2)}}$, to write this coupling in an $SL(2,R)$-covariant family of couplings \cite{Bergshoeff:2006gs}.
}. However, there are other terms in the corresponding S-matrix element at this order. They are the  massless open and closed string poles resulting from the supergravity couplings and the D$_3$-brane couplings $B.f$ and  $C^{(2)}\wedge f$ where $f$ is the world volume gauge field. The combination of all  these terms which are the $\alpha'^0$-order terms of the  disk-level 2-point function \reef{LCS1}, are invariant under the S-duality.

So far  we have discussed only the cases where the S-duality requires one to include the loops and the D-instanton effects to make them invariant under the S-duality transformation on the background. They make the tree-level S-matrix elements to be S-duality invariant. We now discuss the cases where the S-matrix elements are not invariant under the linear S-duality transformation on the quantum states. Hence, the compatibility of disk-level S-matrix elements with the  S-duality requires one to include some other S-matrix elements  at the tree-level. Let us consider, for example, the disk-level 2-point function  of B-field on the world volume of D$_3$-brane.  This amplitude  in the Einstein frame is 
given by \reef{AA} in which the kinematic factor is the  transformation of \reef{LDBI} to the Einstein frame, 
\beqa
K(D_3;B_1,B_2)=e^{-\phi_0}\bigg[e^{-\phi_0}\left(\frac{1}{6}H_{1ijk,a}H_2^{ijk,a}+\frac{1}{3}H_{1abc,i}H_2^{abc,i}-\frac{1}{2}H_{1bci,a}H_2^{bci,a}\right)\bigg]\labell{kin2}
\eeqa
 Since $B\rightarrow - C^{(2)}$ under the S-duality, it is obvious that this S-matrix element is neither invariant under the linear S-duality on the quatum states $B_1,B_2$, nor under nonlinear S-duality on the background. To make it invariant under the linear S-duality of the quantum states, one needs similar   2-point function of the RR two-form.   
 
 The disk-level  2-point function of the RR two-form in the string frame is  given by \reef{A}. The kinematic factor  for the case of D$_3$-brane is \cite{Garousi:2011fc} 
\beqa
K(D_3;C^{(2)}_1,C^{(2)}_2)=e^{2\phi_0}\bigg[\left(\frac{1}{6}F^{(3)}_{1ijk,a}F_2^{(3)ijk,a}+\frac{1}{3}F^{(3)}_{1abc,i}F_2^{(3)abc,i}
-\frac{1}{2}F^{(3)}_{1bci,a}F_2^{(3)bci,a}\right)\bigg]\labell{kin0}
\eeqa
We have normalized the amplitude by $T_3 e^{2\phi_0}$ to make it consistent with the T-duality. To clarify this point, consider implementing T-duality on the world volume of the brane. Under T-duality, the first term, for example, transforms to $F^{(4)}_{1ijky,a}F_2^{(4)ijk}{}_y{}^{,a}$ where $y$ is the Killing index. We need the flat metric $\eta^{yy}$ to contract the $y$ indices. This arises from the nonlinear T-duality on the background dilaton factor which transforms as $e^{2\phi_0}\rightarrow e^{2\phi_0}\eta^{yy}$.

In the Einstein frame, the amplitude is given by \reef{AA} and the following kinematic factor
\beqa
K(D_3;C^{(2)}_1,C^{(2)}_2)=e^{-\phi_0}\bigg[e^{\phi_0}\left(\frac{1}{6}F^{(3)}_{1ijk,a}F_2^{(3)ijk,a}+\frac{1}{3}F^{(3)}_{1abc,i}F_2^{(3)abc,i}
-\frac{1}{2}F^{(3)}_{1bci,a}F_2^{(3)bci,a}\right)\bigg]\labell{kin}
\eeqa
This kinematic factor  is similar to the kinematic factor of two B-fields \reef{kin2}, as predicted by the S-duality. 

The S-duality  predicts even the disk-level S-matrix elements in the presence of constant axion background. Since the RR two-form and the B-field appear as doublet under the S-duality transformation, the following combination is invariant under the S-duality  \cite{Becker:2007zj}:
\beqa
(B_1, C_1^{(2)}){\cal M}\pmatrix{B_2 \cr 
C_2^{(2)}}=e^{-\phi_0}B_1B_2+e^{\phi_0}C_1^{(2)}C_2^{(2)}-e^{\phi_0}C_0(B_1C_2^{(2)}+B_2C_1^{(2)})+e^{\phi_0}C_0^2B_1B_2\nonumber
\eeqa
where $B_1, \, C_1^{(2)}$ and $B_2, \, C_2^{(2)}$ are the polarizations of the external states and the matrix ${\cal M}$ is the following function of the background dilaton and axion:
 \beqa
 {\cal M}=e^{\phi_0}\pmatrix{e^{-2\phi_0}+C_0^2&-C_0 \cr 
-C_0&1}\labell{M}
\eeqa 
 
The consistency of the disk-level 2-point function  of   B-fields  in the zero axion background, \ie equations \reef{AA} and \reef{kin2}, with  the S-duality then predicts   the disk-level 2-point function of the RR two-form in the zero axion background, \ie $ 
e^{\phi_0}C^{(2)}C^{(2)}$, which is given by \reef{AA} and \reef{kin}, and the disk-level 2-point functions $e^{\phi_0}C_0B_1C_2^{(2)}$ and $e^{\phi_0}C_0^2B_1B_2$
in the presence of  non-zero axion background. They all combine into a 2-point function   given by \reef{AA} in which the kinematic factor is
\beqa
K(D_3;\veps_1,\veps_2)=e^{-\phi_0}\bigg[\frac{1}{6}{{\cal F}_1}^T_{ijk,a}\cM\cF_2^{ijk,a}
+\frac{1}{3}{\cF_1}^T_{abc,i}\cM\cF_2^{abc,i}-\frac{1}{2}{\cF_1}^T_{bci,a}\cM\cF_2^{bci,a}\bigg]\labell{kin1}
\eeqa
where $\cF_1=d\cB_1$ and $\cB_1$ is
\beqa
\cB_1=\pmatrix{B_1 \cr 
C_1^{(2)}}\nonumber
\eeqa
Similarly for $\cF_2$. Including the appropriate loops and nonperturbative  effects to the above disk-level S-matrix multiplet, one can  make it  invariant under the nonlinear S-duality on the background.

The disk-level 2-point functions $e^{\phi_0}C_0B_1^{(2)}C_2^{(2)}$ and $e^{\phi_0}C_0^2B_1^{(2)}B_2^{(2)}$ in the presence of non-zero axion background can be calculated with the  zero axion   3-point function $e^{\phi_0}C_3B_1^{(2)}C_2^{(2)}$ and the 4-point function $e^{\phi_0}C_4C_3B_1^{(2)}B_2^{(2)}$, respectively,  in which the axion field in the RR scalar vertex operator is a constant. Note that, in general,  the disk-level 3-point function   and the 4-point function   are much more complicated than the  2-point function. However, when  RR scalar is constant they should be reduced to \reef{AA} with the  kinematic factor \reef{kin1}. It would be interesting to perform these calculations.

The above discussions can be applied for  any other disk-level S-matrix element of D$_3$-brane to find its  S-dual  S-matrix multiplet. One may also extend the above discussions to the sphere-level S-matrix elements because the vacuum  corresponding to the sphere-level is  invariant under the S-duality. In the next section we turn to the cases in which the vacuum is not invariant under the S-duality. 

\section{S-matrix for F$_1$-string and NS$_5$-brane}

We have seen in the previous section how the S-duality invariance of the  zero-axion S-matrix elements on the world volume of D$_3$-brane  may fix the appearance of the axion background in the S-matrix elements. The invariance of the S-matrix elements is related to the fact that the D$_3$-brane is invariant under the S-duality. The D$_1$-brane and D$_5$-brane are  not invariant under the S-duality, hence, one does not expect the S-matrix  elements on the world volume of these branes to be invariant under the S-duality. The S-duality transformation that maps $C^{(2)}\rightarrow B$, transforms D$_1$-brane to F-string and D$_5$-brane to NS$_5$-brane. Hence, in these cases we expect the S-matrix elements to be $SL(2,Z)$-covariant. Using this proposal, in general, one may find the S-matrix elements on the world volume of $(p,q)$-strings and $(p,q)$-5-branes by applying S-duality transformation on the S-matrix element of D$_1$-brane and D$_5$-brane, respectively. This is possible only if one knows the form of the latter S-matrix elements in the presence of axion background. To clarify this, consider the transformation of dilaton-axion field, \ie $\tau=C+ie^{-\phi}$,  under the S-duality
\beqa
\tau\longrightarrow \tau'=\frac{a\tau+b}{c\tau +d}&;& ad-bc=1
\eeqa
If the axion is zero on the left-hand side, then it simplifies to
\beqa
ie^{-\phi}\longrightarrow \tau'=\frac{i e^{-\phi}+(bd+ace^{-2\phi})}{c^2\,e^{-2\phi} +d^2}
\eeqa
which indicates that axion is not zero after duality transformation. So if one begins with the zero-axion S-matrix elements of D$_1$-brane or D$_5$-brane, and applies the above  $SL(2,Z)$ transformation,  then the axion in the transformed S-matrix elements is not zero. On the other hand, since we do not include the axion resulting from the $SL(2,Z)$ transformation of axion, the axion in dual S-matrix does not appear correctly. 

To avoid this difficulty, we use the particular $SL(2,Z)$ transformation which maps D$_1$-brane to F-string and D$_5$-brane to NS$_5$-brane. Under this transformation,
\beqa
\tau\stackrel{s}{\longrightarrow} \tau'=-\frac{1}{\tau }=\frac{ie^{-\phi}-C}{C^2+e^{-2\phi}}\labell{axion}
\eeqa
which indicates if the axion is zero on the left-hand side, it remains zero after duality transformation. So in this section we show how the zero-axion S-matrix elements of F-string and NS$_5$-brane can be read from the zero-axion S-matrix elements of D$_1$-brane and D$_5$-brane, respectively. 

Let us begin with  1-point functions.
 Since D$_1$-string and F$_1$-string couples linearly to the RR two-form and the B-field, respectively, one finds the following transformation on the 1-point function:
\beqa
 T_{D1}\epsilon^{a_0 a_1}C^{(2)}_{a_0a_1}\delta^2(p_1^a)&\stackrel{s}{\longrightarrow}&T_{F1}\epsilon^{a_0a_1}B_{a_0 a_1}\delta^2(p_1^a)\labell{DF}
\eeqa
where the Einstein frame tensions are $T_{D1}=1/(2\pi\alpha'\sqrt{g_s})$ and $T_{F1}=\sqrt{g_s}/(2\pi\alpha')$.  While the first coupling can be confirmed by the disk-level 1-point function in which the RR vertex operator is in $(-1/2,-3/2)$-picture, the second coupling which is a standard coupling, has no such description. There is also the following transformation on the Einstein frame 1-point function of graviton and dilaton \reef{amp3}:
\beqa
 T_{D1}(h_{1a}{}^a-\phi_1)\delta^2(p_1^a)&\stackrel{s}{\longrightarrow}&T_{F1}(h_{1a}{}^a+\phi_1)\delta^2(p_1^a)\labell{DF1}
\eeqa
The graviton couplings on both sides are the standard couplings in the Nambo-Goto action.

 We now  consider the disk-level two-point function of one $C^{(2)}$ and one NSNS state on the world volume of D$_1$-brane. This amplitude in the string frame is given by \reef{A} in which  the appropriate kinematic factor is given in \reef{LTdual}.  In the Einstein frame the amplitude becomes 
\beqa
 A(D_1;\veps_1,\veps_2)&\sim&T_{D1}\alpha'^2K(D_1;\veps_1,\veps_2)\frac{\Gamma(- te^{-\phi_0/2}/4)\Gamma(- se^{-\phi_0/2})}{\Gamma(1-te^{-\phi_0/2}/4- se^{-\phi_0/2})}\delta^{2}(p_1^a+p_2^a)\labell{AAA}
 \eeqa
 where the kinematic factor is
\beqa
K(D_1;C_1^{(2)},h_2)&=&\eps^{a_0a_1}e^{-\phi_0}\bigg[F^{(3)}_{1ia_1j,a}{R_2}^a{}_{a_0}{}^{ij}-F^{(3)}_{1a_0a_1j,i}{\hat{R}}_2^{ij}\bigg]
\eeqa
The amplitude \reef{AAA} is mapped under the S-duality to the  following 2-point function on the world-volume of  the F$_1$-string :
\beqa
 A(F_1;\veps_1,\veps_2)&\sim&T_{F1}\alpha'^2K(F_1;\veps_1,\veps_2)\frac{\Gamma(- te^{\phi_0/2}/4)\Gamma(- se^{\phi_0/2})}{\Gamma(1-te^{\phi_0/2}/4- se^{\phi_0/2})}\delta^{2}(p_1^a+p_2^a)\labell{AAAF}
 \eeqa
 where the kinematic factor is
\beqa
K(F_1;B_1,h_2)&=&\eps^{a_0a_1}e^{\phi_0}\bigg[H_{1ia_1j,a}{R_2}^a{}_{a_0}{}^{ij}-H_{1a_0a_1j,i}{\hat{R}}_2^{ij}\bigg]
\eeqa
 The above transformation is  the extension of the transformations \reef{DF} and \reef{DF1} to 2-point function.

In general, we expect all  2-point function on the world volume of  F$_1$-string at strong coupling to be given by \reef{AAAF} in which the kinematic factor  is related to the kinematic factor of D$_1$-brane by the S-duality transformation, \ie
 \beqa
 K(D_1;\veps_1,\veps_2)&\stackrel{s}{\longrightarrow}&K(F_1;\veps_1,\veps_2)
 \eeqa
 Similarly, one can find  all n-point functions  on the world-volume of F$_1$-string.
 
 As in the case of T-duality, the new S-matrix elements can be found  by using nonlinear S-duality on the background fields and the linear S-duality  on  the quantum fluctuations.
  In fact,    the axion and the dilaton are the only fields which  transform nonlinearly under the S-duality. The background axion is zero in our discussion. The linear S-duality transformation of the quantum state of the axion can be found from \reef{axion} which is 
 \beqa
C&\stackrel{s}{\longrightarrow}&-e^{2\phi_0}C\labell{LC}
\eeqa
where $e^{2\phi_0}$ is the background dilaton factor. Therefore, the axion state in the disk-level n-point function of D$_1$-string is mapped to $-e^{2\phi_0}C$  in the tree-level n-point function of F$1$-string.  

In the string frame, the  amplitude \reef{AAAF} becomes
 \beqa
 A(F_1;\veps_1,\veps_2)&\sim&T_{F1}\alpha'^2K(F_1;\veps_1,\veps_2)\frac{\Gamma(- te^{\phi_0}/4)\Gamma(- se^{\phi_0})}{\Gamma(1-te^{\phi_0}/4- se^{\phi_0})}\delta^{2}(p_1^a+p_2^a)\labell{ACC}
 \eeqa
where the string tension is $T_{F1}=1/(2\pi\alpha')$. The gamma functions  represent the $s$- and $t$-channels. The poles in the $t$-channel are at $\frac{g_st}{2}=0,2,4,\cdots$  and the poles in the $s$-channel are at $\frac{g_ss}{2}=0,\frac{1}{2},1,\cdots$. These two channels are similar to the $s$- and $t$-channels of the disk-level scattering from D-branes \cite{Garousi:1996ad}. In that case the poles in the $t$-channel are at $\frac{t}{2}=0,2,4,\cdots$, and the poles in the $s$-channel are at $\frac{s}{2}=0,\frac{1}{2},1,\cdots$. In terms of the  tension of the intermediate string, the $t$-channel poles in the D-brane amplitude are at $-(p_1+p_2)^2/(2\pi T_{F1})=0,2,4,\cdots$, whereas the $t$-channel poles in the F$_1$-string  amplitude are at $-(p_1+p_2)^2/(2\pi T_{D1})=0,2,4,\cdots$. Hence, the extra factor of $g_s$ in  the  F$_1$-string amplitude \reef{ACC} dictates that the intermediate string  is  D$_1$-string, as expected.  


The magnetic dual of the  transformation \reef{DF} is
\beqa
 T_{D5}\epsilon^{a_0\cdots a_5}C^{(6)}_{a_0\cdots a_5}\delta^6(p_1^a)&\stackrel{s}{\longrightarrow}&T_{NS5}\epsilon^{a_0\cdots a_5}B_{a_0 \cdots a_5}\delta^6(p_1^a)\labell{DFM}
\eeqa
The Einstein frame tensions are $T_{D5}=\sqrt{g_s}/[4\pi^2(2\pi\alpha')^3]$ and $T_{NS5}=1/[4\pi^2(2\pi\alpha')^3\sqrt{g_s}]$.  Repeating the same steps as we have done for  \reef{DF}, one finds the following 2-point function for the NS$_5$-brane at strong coupling in the Einstein frame:
\beqa
 A(NS_5;\veps_1,\veps_2)&\sim&T_{NS5}\alpha'^2K(NS_5;\veps_1,\veps_2)\frac{\Gamma(- te^{\phi_0/2}/4)\Gamma(- se^{\phi_0/2})}{\Gamma(1-te^{\phi_0/2}/4- se^{\phi_0/2})}\delta^{6}(p_1^a+p_2^a)\labell{ACCM}
 \eeqa
 where the kinematic factor $K(NS_5;\veps_1,\veps_2)$ is related to the kinematic factor of D$_5$-brane by the S-duality transformation, \ie
 \beqa
 K(D_5;\veps_1,\veps_2)&\stackrel{s}{\longrightarrow}&K(NS_5;\veps_1,\veps_2)
 \eeqa
 where $ K(D_5;\veps_1,\veps_2)$ is the kinematic factor of the D$_5$-brane in the Einstein frame.  We expect in a similar way   all other  S-matrix elements  can be found.
 
We have seen that the S-matrix elements are invariant/covariant under the global S- and T-dualities. String theory is also invariant under  the global supersymmtry. Hence,  one  expects the S-matrix elements to be invariant under the supersymmetry as well. In this case we call the S-matrix elements which are interconnected by the supersymmetry, a  supersymmetric S-matrix multiple. When  the  supersymmetry transformations \cite{Schwarz:1983wa} are used to transform all the bosonic and the fermionic components of  the multiplet, the supersymmetry parameter $\epsilon$  must   be canceled. In other words,  the supersymmetric S-matrix multiplet should satisfy the Ward identity associated with the global supersymmetry transformations.  
It has been shown in \cite{Garousi:1996ad}  the disk-level 2-point functions satisfy the Ward identities corresponding to all the gauge symmetries. It would be interesting to  show  that they satisfy the Ward identity corresponding to the global supersymmetry as well. 

The  F$_1$-string/NS$_5$-brane S-matrix elements that we have found are the S-dual of the disk-level D$_1$-string/D$_5$-brane S-matrix elements, hence,  they are valid at strong coupling. Since the S-matrix elements are invariant under the supersymmetry, one expects the above F$_1$-string/NS$_5$-brane S-matrix elements to be valid for any coupling. However, the  loops and the non-perturbative effects in these S-matrix elements which are the S-dual of the corresponding effects in D$_1$-string/D$_5$-brane S-matrix elements, have non-zero contributions at arbitrary coupling.  We have seen  in the previous section that these  effects have no contribution in $O(\alpha'^0)$-order terms. Therefore, the $O(\alpha'^0)$-order terms of the above S-matrix elements are valid at any arbitrary coupling.
 
\subsection{Massless poles}

The above  amplitudes for F$1$-string and NS$_5$-brane indicate that there are massless poles in both open and closed D$_1$-string channels. Let us examine these poles. The scattering amplitude of one RR scalar (the axion quantum fluctuation) and one RR two-form on the world-volume of F$_1$-string  is given by \reef{AAAF}. At the leading order in $\alpha'$ it is
\beqa
A(F_1;C_1,C_2^{(2)})&\sim&T_{F1}\alpha'^2e^{3\phi_0}\epsilon^{a_0a_1}\bigg[F^{(1)}_{1i,a}F^{(3)}_{2a_0a_1}{}^{a,i}-F^{(1)}_{1a,i}F^{(3)}_{2a_0a_1}{}^{i,a}\bigg]\left(-\frac{1}{e^{\phi_0}st}+\cdots\right)\labell{AF1L}
\eeqa
where the metric is the Einstein metric. The corresponding amplitude for D$_1$-string is
\beqa
A(D_1;C_1,B_2)&\sim&T_{D1}\alpha'^2e^{-\phi_0}\epsilon^{a_0a_1}\bigg[F^{(1)}_{1i,a}H_{2a_0a_1}{}^{a,i}-F^{(1)}_{1a,i}H_{2a_0a_1}{}^{i,a}\bigg]\left(-\frac{1}{e^{-\phi_0}st}+\cdots\right)\nonumber
\eeqa
One can easily observe that the F$_1$-string amplitude is the transformation of the D$_1$-string amplitude under the S-duality.

Now we are going to reproduce the above amplitudes in effective field theory. Using the following standard coupling in the type IIB supergravity in the Einstein frame \cite{Becker:2007zj}:
\beqa
\int d^{10}x \sqrt{-g}{\cal H}_{\mu\nu\rho}^T{\cal M}{\cal H}^{\mu\nu\rho}\labell{FCH}
\eeqa
where the matrix ${\cal M}$ is given in \reef{M} and 
\beqa
{\cal H}_{\mu\nu\rho}&=&\pmatrix{H_{\mu\nu\rho} \cr 
F^{(3)}_{\mu\nu\rho}},
\eeqa
and the standard linear coupling of the B-field to F$_1$-string \reef{DF}, one can calculate the massless closed D-string pole in the scattering 
\beqa
C+{\rm F_1\!\!-\!string}&\longrightarrow &C^{(2)}+{\rm F_1\!\!-\!string}\labell{22}
\eeqa
The Feynman amplitude becomes
\beqa
{\cal A}_t(F_1)&\sim&T_{F1}\frac{F_1^{(1)\mu}F^{(3)}_{2\mu a b}\epsilon^{ab}}{t}e^{2\phi_0}\labell{AF12}
\eeqa
On the other hand, the supergravity coupling \reef{FCH} and the  linear coupling of the RR two-form to D$_1$-string \reef{DF} can be used to  calculate the massless closed string pole in the following scattering: 
\beqa
C+{\rm D_1\!\!-\!string}&\longrightarrow &B^{(2)}+{\rm D_1\!\!-\!string}\labell{24}
\eeqa
The Feynman amplitude in this case becomes
\beqa
{\cal A}_t(D_1)&\sim&T_{D1}\frac{F_1^{(1)\mu}H_{2\mu a b}\epsilon^{ab}}{t}\labell{AD1}
\eeqa
Comparing this amplitude with \reef{AF12} and using the linear transformation of the axion \reef{LC}, one finds that the massless closed string poles are related to each other by the S-duality,
which is consistent with our proposal for the string amplitude \reef{AF1L}.

The massless open string pole in the scattering \reef{24} can be calculated by using the standard brane couplings in the Einstein frame: $T_{D1}B_{ab}f^{ab}e^{\phi_0/2}$, $T_{D1}\epsilon^{ab}f_{ab}C$ and $T_{D1}f_{ab}f^{ab}e^{\phi_0/2}$. The Feynman amplitude becomes
\beqa
{\cal A}_s(D_1)&\sim&T_{D1}\frac{\epsilon^{ab}F^{(1)}_{1a}B_{2bc}{}^{,c}}{s}\labell{AsD}
\eeqa
On the other hand, the massless open D-string pole in \reef{22} can be found by assuming the brane couplings $T_{F1}C_{ab}\hat{f}^{ab}e^{-\phi_0/2}$, $T_{F1}\epsilon^{ab}\hat{f}_{ab}Ce^{2\phi_0}$ and $T_{F1}\hat{f}_{ab}\hat{f}^{ab}e^{-\phi_0/2}$ in the Einstein frame where $\hat{f}_{ab}$ is the S-dual of the gauge field $f_{ab}$. The Feynman amplitude becomes
\beqa
{\cal A}_s(F_1)&\sim&T_{F1}\frac{\epsilon^{ab}F^{(1)}_{1a}C_{2bc}{}^{,c}}{s}e^{2\phi_0}\labell{AsF}
\eeqa
Comparing \reef{AsD} with \reef{AsF}, one again finds that the massless open string poles are related to each other by S-duality, which is consistent  with the proposal for the string amplitude \reef{AF1L}.

One can find the contact terms at order $O(\alpha'^0)$ by subtracting the above massless poles from the $\alpha'$-expansion of the tree-level 2-point function. In the case of D$_1$-string, one finds
\beqa
A(D_1;C_1,B_2)-{\cal A}_t(D_1)-{\cal A}_s(D_1)&\sim&T_{D1}C_1B_{2ab}\epsilon^{ab}+O(\alpha'^2)
\eeqa
which is a standard term in the Chern-Simons part of the D$_1$-string action. Similar calculation for F$_1$-string gives
\beqa
A(F_1;C_1,C_2^{(2)})-{\cal A}_t(F_1)-{\cal A}_s(F_1)&\sim&T_{F1}C_1C_{2ab}\epsilon^{ab}e^{2\phi_0}+O(\alpha'^2)
\eeqa
This is a coupling in the world volume of F$_1$-string. One may extend the above calculations to the other scattering amplitudes and find other  couplings in the effective action of F$_1$-string/NS$_5$-brane. In the next section we discuss these couplings.

\section{F$_1$-string and NS$_5$-brane effective actions}

The dynamics of the D-branes of type II superstring theories is well-approximated by the effective world-volume field theory  which consists of the  Dirac-Born-Infeld (DBI) and the Chern-Simons (CS) actions. 
The DBI action  describes the dynamics of the brane in the presence of  NS-NS background fields, which  can be found by requiring its  consistency with the  nonlinear T-duality \cite{Leigh:1989jq,Bachas:1995kx}. On the other hand,  the CS part describes the coupling of D-branes to the R-R potential \cite{Polchinski:1995mt,Douglas:1995bn}. These actions in the string frame for 
  D$_1$-brane  and  D$_5$-brane are\footnote{We are using the convention in which the asymptotic value of the dilaton is zero. In this convention the D-brane tension and the D-brane charge are identical \cite{Myers:1999ps}.} 
\beqa
S_{D1}&=&-T_{D1}\int d^2xe^{-\phi}\sqrt{-\det(g_{ab}+B_{ab})}+T_{D1}\int [C^{(2)}+CB]\labell{DBICS}\\
S_{D5}&=&-T_{D5}\int d^6xe^{-\phi}\sqrt{-\det(g_{ab}+B_{ab})}\nonumber\\
&&+T_{D5}\int [C^{(6)}+C^{(4)}\wedge B+\frac{1}{2}C^{(2)}\wedge B\wedge B+\frac{1}{3!}CB\wedge B\wedge B]\nonumber
\eeqa
All the closed string fields in the actions are pull-back of the bulk fields onto  the world-volume of branes. The abelian gauge field can be added to the actions as $B\rightarrow B+2\pi\alpha'f$. This makes the action to be invariant under the B-field gauge transformation. These actions can be  naturally extended to the nonabelian case by using the symmetric trace prescription  \cite{Tseytlin:1997csa,Tseytlin:1999dj}, and by including the Myers terms \cite{Myers:1999ps}.  The above actions can be confirmed by  the contact terms of the D-brane  S-matrix elements  at order $O(\alpha'^0)$.

We have seen in section 3, how to find the S-matrix elements of F$_1$-string and NS$_5$-brane at zero axion background by applying the particular S-duality transformation \reef{axion} on the disk-level S-matrix element of D$_1$-string and D$_5$-brane, respectively. If one knew the disk-level S-matrix elements in non-zero axion background, then the S-duality would produce the  S-matrix elements of F$_1$-string and NS$_5$-brane at non-zero axion background. In that case also the contact terms of D$_1$-string/D$_5$-brane at order $O(\alpha'^0)$ would be mapped to the contact terms of F$_1$-string/NS$_5$-brane at order $O(\alpha'^0)$. On the other hand, the contact terms of D$_1$-string/D$_5$-brane produce the effective action \reef{DBICS}. Hence, the contact terms of F$_1$-string/NS$_5$-brane at order $O(\alpha'^0)$ should produce  effective actions which are related to \reef{DBICS} by the S-duality.

To apply the S-duality \reef{axion} on \reef{DBICS}, it is better to first write them in the Einstein frame. There is no metric in the Chern-Sioms parts, so they remain unchanged. The DBI parts in the Einstein frame become
\beqa
S_{D1}(DBI)&=&-T_{D1}\int d^2xe^{-\phi/2}\sqrt{-\det(g_{ab}+e^{-\phi/2}B_{ab})}\labell{DBICSE}\\
S_{D5}(DBI)&=&-T_{D5}\int d^6xe^{\phi/2}\sqrt{-\det(g_{ab}+e^{-\phi/2}B_{ab})}\nonumber
\eeqa
The tensions  in the Einstein frame are
\beqa
T_{D1}=1/(2\pi\alpha'\sqrt{g_s})&;&T_{D5}=\sqrt{g_s}/[4\pi^2(2\pi\alpha')^3]
\eeqa
Note that if we had used the conversion in which the asymptotic value of the dilaton in \reef{DBICS} were non-zero, then the string coupling $g_s$ would not appear in the above tensions. 

The particular S-duality transformation that maps  D$_1$-string to F$_1$-string and D$_5$-brane to NS$_5$-brane is 
\beqa
C^{(2)}\stackrel{s}{\longrightarrow} B&;&B\stackrel{s}{\longrightarrow} -C^{(2)}\nonumber\\
C^{(6)}\stackrel{s}{\longrightarrow} B^{(6)}&;&B^{(6)}\stackrel{s}{\longrightarrow} -C^{(6)}\nonumber\\
e^{-\phi}\stackrel{s}{\longrightarrow} \frac{1}{C^2e^{\phi}+e^{-\phi}}&;&C\stackrel{s}{\longrightarrow} -\frac{C}{C^2+e^{-2\phi}}\nonumber\\
g_{\mu\nu}\stackrel{s}{\longrightarrow} g_{\mu\nu}&;&C^{(4)}\stackrel{s}{\longrightarrow} C^{(4)}
\eeqa
 Using these transformation,  one finds that the actions \reef{DBICSE}  are mapped to the following actions:
\beqa
S_{F1}&\!\!\!\!\!\!\!\!=\!\!\!\!\!\!\!\!&T_{F1}\int\bigg[- d^2x\left(C^2e^{\phi}+e^{-\phi}\right)^{-1/2}\sqrt{-\det\left(g_{ab}-\frac{C_{ab}}{\left(C^2e^{\phi}+e^{-\phi}\right)^{1/2}}\right)}+B+\frac{CC^{(2)}}{C^2+e^{-2\phi}}\bigg]\nonumber\\
S_{NS5}&\!\!\!\!\!\!\!\!=\!\!\!\!\!\!\!\!&T_{NS5}\int\bigg[- d^6x\left(C^2e^{\phi}+e^{-\phi}\right)^{1/2}\sqrt{-\det\left(g_{ab}-\frac{C_{ab}}{\left(C^2e^{\phi}+e^{-\phi}\right)^{1/2}}\right)}\labell{final}\\
&&\qquad\qquad\qquad+B^{(6)}-C^{(4)}\wedge C^{(2)}+\frac{1}{2}B\wedge C^{(2)}\wedge C^{(2)}+\frac{1}{3!}\frac{CC^{(2)}\wedge C^{(2)}\wedge C^{(2)}}{C^2+e^{-2\phi}}\bigg]\nonumber
\eeqa
where the Einstein frame tensions are
\beqa
T_{F1}=\frac{1}{2\pi\alpha'}\left(C_0^2e^{\phi_0}+e^{-\phi_0}\right)^{-1/2}&;&T_{NS5}=\frac{1}{4\pi^2(2\pi\alpha')^3}\left(C_0^2e^{\phi_0}+e^{-\phi_0}\right)^{1/2}
\eeqa
 The closed string fields in \reef{final} are pull-back of the bulk fields onto  the world-volume of branes.  The abelian gauge field are  added to the actions by the replacement $C^{(2)}\rightarrow C^{(2)}+2\pi\alpha'\hat{f}$. Then the above actions are invariant under the RR two-form gauge transformation. The quadratic couplings at order $O(\alpha'^0)$ that considered in the previous section are consistent with the above actions. As we argued before in section 3, the $O(\alpha'^0)$-order terms of the S-matrix elements receive no loops or nonperturbative corrections, hence, the above actions are expected to be valid for any string coupling.
 
 The gauge field $\hat{f}_{ab}$ and the transverse scalar fields in the definition of the pull-back operation in \reef{final} which are the transformation of the corresponding fields in \reef{DBICS} under the S-duality,  are the massless open D-string excitation of $F_1$-string/NS$_5$-brane. The transformation $f\stackrel{s}{\longrightarrow} -\hat{f}$ under the S-duality has been considered in \cite{Townsend:1997kr} in proposing an S-dual action for superstring.  
The above actions can be  extended to the nonabelian case by using the symmetric trace prescription, and by including the Myers terms in which $C^{(2)}, C^{(6)}$ are replaced by $B^{(2)}, B^{(6)}$. 


Finally, let us compare our results with the results in \cite{Bergshoeff:2006gs}. An $SL(2,R)$-covariant  action for all D$_p$-branes of type IIB string theory   which  is based on the assumption that the Chern-Simons part is gauge invariant, has been proposed in \cite{Cederwall:1997ab,Bergshoeff:2006gs}. The gauge field in the action of D$_1$-brane has been integrated out in  \cite{Bergshoeff:2006gs}, and hence the  action for F$_1$-string  in \cite{Bergshoeff:2006gs} has no RR two-form. Doing the same thing here, one would find the same result as in \cite{Bergshoeff:2006gs}. The action \reef{final} for NS$_5$-brane is consistent with the $SL(2,R)$-covariant action for $(p,q)$-5-brane proposed in \cite{Bergshoeff:2006gs} (see equation (3.5) in \cite{Bergshoeff:2006gs} for the special charge of $q=\pmatrix{1 \cr 
0}$). Note that in the convention  \cite{Bergshoeff:2006gs}, the asymptotic value of the dilaton is non-zero, hence the tension of D$_5$-brane and NS$_5$-brane are  constant that depends only on $\alpha'$ which have been dropped in \cite{Bergshoeff:2006gs}. 

We have seen in section 2 that the S-matrix elements on the world volume of D$_3$-brane can be combined into  the $SL(2,Z)$-invariant multiplets. In particular, the $O(\alpha'^0)$-order terms of the disk-level S-matrix elements can be combined into  the $SL(2,Z)$-invariant multiplets without adding the loops and the non-perturbative effects. These $O(\alpha'^0)$-order terms include  massless poles and contact terms. Hence, the contact terms which produce the DBI and Chern-Simons actions of the D$_3$-brane, are not invariant under the S-duality. In this respect, it  is similar to the discussions in \cite{Gibbons:1995ap,Tseytlin:1996it,Green:1996qg} that show even tough the D$_3$-brane action is not invariant under the S-duality, however, the equations of motion resulting from this action and the supergravity are invariant under the S-duality. 

Alternatively,  one may use the $SL(2,R)$-doublets
$
\tq=\pmatrix{1 \cr 
0}$, $ q=\pmatrix{0 \cr 
-1}
$
and rotate them at the same time that rotate the world volume and bulk polarization tensors, to write  the S-matrix elements on the world volume of D$_3$-brane  in an $SL(2,R)$-covariant family of S-matrix elements. In this case, the intermediate string propagating on the world volume of D$_3$-brane would be the $(p,q)$-string. This would be unlike the $SL(2,Z)$-invariant case that the intermediate string  is always $F_1$-string. The massless poles and the contact terms of the $SL(2,R)$-covariant  S-matrix elements at order  $O(\alpha'^0)$ should be separately covariant. The contact terms should then be reproduced by the $SL(2,R)$-covariant action proposed in  \cite{Bergshoeff:2006gs}.

 {\bf Acknowledgments}:  I would like to thank Ashok Sen for useful  discussions. This work is supported by Ferdowsi University of Mashhad under grant 2/17837-1390/03/24.



\end{document}